\documentclass[preprint]{aastex}

\begin{document}

\newcommand{\kms}{\,km\,s$^{-1}$}     

\title{FUSE Observations of Atomic Abundances and
Molecular Hydrogen in the Leading Arm of the Magellanic Stream}

\author{Kenneth. R. Sembach\altaffilmark{1}, 
	J. Christopher Howk\altaffilmark{1}, 
        Blair D. Savage\altaffilmark{2}, 
	J. Michael Shull\altaffilmark{3}}
\altaffiltext{1}{Department of Physics \& Astronomy, The Johns 
	Hopkins University, Baltimore, MD  21218}
\altaffiltext{2}{Department of Astronomy, University of Wisconsin, Madison, 
	WI  53706}
\altaffiltext{3}{CASA and JILA, 
Department of Astrophysical and Planetary Sciences,
University of Colorado, Boulder, CO  80309}

\slugcomment{To appear in the Astronomical Journal.}

\begin{abstract}
We present {\it Far Ultraviolet Spectroscopic Explorer} observations of 
the atomic and molecular absorption in high velocity cloud 
HVC\,287.5+22.5+240, which lies in front of the ultraviolet-bright 
nucleus of the Seyfert 1 galaxy NGC\,3783.    
We detect H$_2$, \ion{N}{1},
\ion{N}{2}, \ion{Si}{2}, and \ion{Fe}{2} and set limits
on the amount of absorption
due to \ion{P}{3}, \ion{Ar}{1}, and \ion{Fe}{3}.  
We extend the earlier metallicity and dust-depletion measurements made by
Lu and collaborators by examining the relative gas-phase abundances of 
Si, P, S, and Fe.   Corrections to the derived gas-phase abundances due to
ionized gas in the HVC are small ($\lesssim15$\%).  
The HVC has a metallicity of 0.2--0.4 solar,
similar to that of the Small Magellanic Cloud. 
The relative abundance pattern for the elements studied resembles that of 
warm gas in the SMC, which
supports the idea that this HVC is part of the tidally stripped 
Leading Arm of the Magellanic Stream.
The abundance pattern implies that the HVC contains dust grains that
have been processed significantly; it is likely that the grain mantles 
have been modified or stripped back to expose the grain cores.
We have identified more than 30 lines of 
H$_2$ arising in the HVC from rotational levels J\,=\,0 to J\,=\,3.
Synthetic spectra and a curve-of-growth fit to these lines with b = 12 \kms\
indicate that 
log\,N(H$_2$) = 16.80$\pm0.10$ and 
f$_{H_2}$\,=\,2N(H$_2$)/[N(\ion{H}{1})+2N(H$_2$)] 
= $1.6\times10^{-3}$.  
A two-component temperature distribution 
is necessary to explain the observed populations of the H$_2$ rotational
levels.  We find $T_{01} = 133\pm^{37}_{21}$\,K, and $T_{23} = 
241\pm^{20}_{17}$\,K, indicating that the 
conditions in the molecular gas are more similar to those found for diffuse
molecular clouds in the Galactic halo than to those for molecular clouds
in the Galactic disk.  
From an analysis of the $J = 2$ and $J = 3$ populations, we find an 
absorption rate 
(at 1000\,\AA) of $\beta_{\rm uv} <0.1$ times the average value in the solar 
neighborhood.
The presence of 
molecular gas in the HVC  requires that either the
H$_2$ formed in situ or that molecules formed within the SMC 
survived tidal stripping.  We favor the latter possibility because of the 
long H$_2$ formation time ($\sim10^8$ years) derived for this HVC.

\end{abstract}


\section{Introduction}
The spatial distribution and velocities of neutral interstellar clouds 
moving at high velocities ($|V_{LSR}| > 100$ \kms) are well characterized 
down to \ion{H}{1} column density levels of $\sim10^{18}$ cm$^{-2}$, but 
the fundamental properties of most high velocity clouds 
(HVCs) remain highly uncertain because of limited information 
about their distances,
physical properties, elemental abundances, and dust content
(see Wakker \& van~Woerden 1997 for a review).  However, for HVC 
Complex~C and the Magellanic Stream, ultraviolet absorption-line 
spectroscopy has yielded important information about the elemental 
composition of the gas and has shed new light on the origins of these 
HVCs. HVC Complex C lies at a distance d $\gtrsim $10 kpc, has a 
metallicity of $\sim$0.1--0.3 solar, and apparently contains little dust 
(Wakker et al. 1999; Gibson et al. 2000b; Murphy et al. 2000),
which indicates that it is not material ejected into the halo as
a consequence of supernovae in the Galactic disk.  Rather, it is probably 
metal-poor infalling gas that may be mixing with
circulating Galactic gas at large distances from the Galactic plane. 
In contrast, \ion{H}{1} HVC\,287.5+22.5+240 has a metallicity 
(S/H = $0.25\pm0.07$ solar,
Lu et al. 1994a), dust content (Fe/H $\approx$ 0.033 solar, Lu et al. 1998),
and velocity that suggest it is associated with the Magellanic 
Stream, an extended ($\sim$100\degr)
 trail of gas pulled out of 
the Magellanic Clouds during a recent close encounter with the Galaxy
(Wannier \& Wrixon 1972; Mathewson, Cleary, \& Murray 1974). 

In this article we report on high-resolution ($\lambda/\Delta\lambda \approx
20,000$) far-ultraviolet (far-UV) absorption-line data 
obtained for HVC\,287.5+22.5+240 with the {\it Far Ultraviolet Spectroscopic 
Explorer} (FUSE), a powerful new observatory for studying the molecular,
atomic, and ionized gases in HVCs and other interstellar/intergalactic clouds
(see Sembach 1999; Oegerle et al. 2000;
Sembach et al. 2000b; Shull et al. 2000a).
 A map of the high velocity 
\ion{H}{1} in the general direction of NGC\,3783 is shown in Figure~1.
A fortuitous alignment of HVC\,287.5+22.5+240 and the 
nucleus of NGC\,3783 ($l = 287.46\degr$, $b = +22.95\degr$), a 
Seyfert 1 galaxy with a bright ultraviolet continuum, makes it possible to 
study the HVC in absorption below 1200\,\AA\ with FUSE.  The HVC has been 
studied previously in \ion{H}{1} 21\,cm radio emission (Mathewson et al. 
1974; Hulsbosch 1975; Morras \& Bajaja 1983; Wakker et al. 1999) 
and at longer UV
wavelengths with the {\it Hubble Space Telescope} (HST) by Lu et al. 
(1994a, 1998)\footnotemark.  
\footnotetext{This HVC is identified as HVC\,187 in the Wakker \& van~Woerden
(1991) catalog of high velocity clouds.}
The HVC position and velocity
of +240 \kms\ are consistent with gas arising in the Leading Arm 
of the Magellanic Stream.  The Leading Arm is predicted by N-body simulations 
tracing the tidal interactions of the Clouds and the Galaxy
(Gardiner \& Noguchi 1996) and has been identified in \ion{H}{1} 21\,cm 
emission (Putman et al. 1998; Putman 2000). 

The FUSE spectrum of NGC\,3783 contains unique signatures of the absorption 
within HVC 287.5+22.5+240, including the first detections of H$_2$ in the 
Magellanic Stream.  The spectrum also allows for 
a quantitative estimate of the ionized gas 
content of the HVC to be made, 
which is crucial for determining accurate abundances.  
This is particularly important in light of recent FUSE results 
showing that many HVCs have a collisionally ionized component seen in
\ion{O}{6} absorption.  This highly ionized gas probably occurs at either the 
interfaces or mixing layers between neutral HVC gas and a hot Galactic halo or
within cooling regions of hot gas clouds as they are accreted onto the Galaxy 
(Sembach et al. 2000b). 

This paper is organized as follows:  Section 2 contains 
a summary of the FUSE observations and data processing.  In \S\,3 we 
describe 
the HVC absorption and our measurements of various species.  In \S\,4 we
consider the atomic and molecular components of the HVC and provide a short
description of the intermediate velocity gas along the sight line.
In \S\,5 we discuss the FUSE results and their implications for the origin and
evolution of HVC\,287.5+22.5+240 and the Magellanic Stream.

\section{Observations and Data Reduction}

The FUSE data for this investigation were obtained on 2 February 2000 and are
stored in the FUSE archive at the Multi-Mission Archive at
the Space Telescope Science Institute under the identifications P10133001 
through P10133013.
FUSE contains two microchannel plate detectors and four co-aligned optical 
channels, with two channels covering shorter wavelengths (SiC; 905--1100\,\AA) 
and two channels optimized for longer wavelengths (Al+LiF; 1000--1187\,\AA).
NGC\,3783 was 
centered in the large (LWRS; $30\arcsec\times30\arcsec$) aperture of 
the LiF1 channel of 
detector 1 for 13 exposures totaling 37 ksec.   (This is the channel used 
for guiding.) All four channels were reasonably well co-aligned throughout 
the observation, but the exposure times on the SiC2 and LiF2 channels 
were limited to 19.5 ksec by the shutdown of detector 2 during exposures 
9--13.
Most of the FUSE band is covered by at least two channels, except for a 
few small spectral intervals ($\lambda < 916$\,\AA, 
$\lambda > 1182$\,\AA).  We restrict our attention in this article to 
the data from the LiF1 channel and use the LiF2 
data to check for fixed-pattern noise and artifacts.  At wavelengths where the 
LiF channels do not provide coverage because of gaps between the detector 
segments
($\sim1082-1095$\,\AA), we use data 
from the SiC2 channel.   A more complete description of the instrument is 
given by Moos et al. (2000), and the on-orbit performance is discussed by 
Sahnow et al. (2000). 

The raw time-tagged photon event lists for each exposure were run through the 
standard FUSE calibration pipeline (CALFUSE v1.7.5) available at the Johns 
Hopkins University as of June 2000.  We screened the lists for valid data 
with the usual 
constraints imposed for Earth limb avoidance and passage through the South 
Atlantic Anomaly.  We applied corrections for geometric distortions of 
the spectra on the detectors, spectral motions on 
the detectors induced by thermal changes, and Doppler shifts cause by orbital 
motion.    Suitable astigmatism and flatfield calibrations were 
not available at the time of this study.  Therefore, we required that any 
absorption features analyzed be 
present in the data from at least two channels.
Scattered light and detector backgrounds are negligible at the flux levels 
of the absorption lines considered in this study.  The processed LiF1A data 
were rebinned to a sampling interval of 
$\approx8$ \kms\ per pixel since the extracted data are oversampled.  The data 
have a nominal (astigmatic) spectral resolution of $\sim20$ \kms\ 
(2.5 pixel FWHM) and S/N = 10 to 15 per resolution element. The relative 
wavelength solution is accurate to 
approximately 6 \kms\ (1 $\sigma$) when averaged over an entire channel, but 
there are a few wavelengths where the solution may fail by as much as 10--15 
\kms\ over small (5--10\,\AA) intervals.  We set the zero point of the 
wavelength scale by comparing the low velocity interstellar lines 
observed in the FUSE bandpass to the longer wavelength
\ion{S}{2} and \ion{Fe}{2} lines observed with the HST by Lu et al. (1998).
This zero point also provides a consistent match between the velocities of the 
low ion absorption line profiles and the \ion{H}{1} 21\,cm emission along the 
sight line.

Selected regions of the FUSE spectrum of NGC\,3783 are shown in Figure~2.  
Prominent interstellar \ion{Fe}{2} and \ion{Ar}{1} lines are labeled above 
the spectrum.  Absorption features due to H$_2$ in both the low 
velocity interstellar gas and the HVC (at +240 \kms) are 
indicated with tick marks.  The regular spacing of the H$_2$ lines
in the Lyman series (0--0), (3--0), and (4--0) bands makes identification
of the HVC features unambiguous. The continuum, emission, 
and absorption features intrinsic to NGC\,3783 in other portions of  
the FUSE spectrum
will be described by Kaiser et al. 
(2001) and will not be discussed here, other than to note possible 
blending with the interstellar lines when appropriate.

\section{Interstellar Measurements}

The NGC\,3783 spectrum is complex, with absorption lines in the HVC 
overlapping absorption features
from the intrinsic systems associated with NGC\,3783 and the interstellar
medium (ISM) of the Milky Way.  For cases where the blending is not too
severe, we determined continua for the interstellar and HVC lines in the
spectrum 
by fitting low-order ($<5$) Legendre polynomials to nearby 
($\pm500$ \kms) regions of the 
AGN continuum.  Equivalent widths for atomic and molecular lines measured in 
the high velocity gas are listed in Table~1.  The equivalent widths and 
errors were derived following the prescription outlined by Sembach 
\& Savage (1992).   Species detected in the HVC include:  H$_2$, \ion{N}{1},
\ion{Si}{2}, and \ion{Fe}{2}.  Additional atomic species 
(e.g., \ion{H}{1}, \ion{C}{2}, 
\ion{C}{3}, and \ion{O}{1}) are present but are severely blended with other 
absorption features or fall in wavelength regions of low data quality.  

In Table~1 we have included values for the \ion{N}{2} $\lambda1083.99$ line 
and three lines of H$_2$ in the Lyman (0--0) vibrational band near 1092\,\AA.
We measured these lines using the available SiC2 data since they fall in the 
LiF wavelength gap.  The SiC2 
data are of lower quality, so these values have larger errors.  

A selection of continuum normalized line profiles is shown in Figure~3.  
The stronger atomic lines reveal absorption at low 
velocities (V$_{LSR} \sim 0$ \kms), at intermediate velocities 
(V$_{LSR} \sim +60$ \kms) and at high velocities 
(V$_{LSR} \sim +240$ \kms).  These velocity components correspond to local gas 
in the disk of the Galaxy, intermediate velocity gas in the low halo 
(z $\sim$ 5 kpc, assuming
corotation with underlying disk gas), and high velocity gas in the HVC.
Column densities for all of the atomic lines listed in Table~1 
were calculated by direct integration of the apparent optical depth profiles
(Savage \& Sembach 1991) defined by:

\begin{equation}
N_a = \int N_a(v) dv
= \frac{3.768\times10^{14}}{f\lambda} \int \tau_a(v) dv~~~~(cm^{-2})
\end{equation}

\noindent
where $\tau_a(v)$ is the apparent optical depth of the line (equal to the 
natural logarithm of the estimated continuum divided by the observed intensity)
at velocity $v$ (in \kms), $f$ is the oscillator strength of the 
line, and $\lambda$ is the wavelength of the line (in \AA).  
For the lines considered here, $N \approx N_a$ since comparisons of 
the \ion{Fe}{2} $\lambda\lambda2344.214, 2374.461$ lines observed by 
the GHRS show that there are no significant unresolved saturated structures 
within the atomic gas (see Figure~3 in Lu et al. 1998).  
The GHRS data have a resolution that is better than
the FUSE data ($\sim13$ \kms\ versus $\sim20$ \kms), but the 
HVC lines observed by FUSE are weaker
than the longer wavelength \ion{Fe}{2} lines (see Figure~3).

\section{Results for HVC\,287.5+22.5+240}
\subsection{Atomic Species}

Several atomic species are detected in HVC\,287.5+22.5+240.  
Detections free of contamination by other absorption features are available 
for the lines of \ion{N}{1} $\lambda1134.415$, 
\ion{Si}{2} $\lambda1020.699$, and 
\ion{Fe}{2} $\lambda1144.938$.  
There is a marginal detection of \ion{N}{2} derived from the 1083.990\,\AA\ 
line, but this detection is of lower quality since it was obtained using 
SiC2 data.  Upper limits can be derived for \ion{Fe}{2} 
$\lambda\lambda 1143.226$, 1125.448; 
\ion{Fe}{3} $\lambda1122.526$; \ion{P}{2} $\lambda1152.818$; and  \ion{P}{3} 
$\lambda998.000$.  An additional upper limit can also be derived for 
the \ion{Ar}{1} $\lambda1048.220$ line, which occurs on the blue wing
of the Lyman (4--0) R(0) line at 1049.367\,\AA\ (see Figure~2).

The combination of the \ion{Fe}{2} and \ion{Si}{2} 
detections with the limits available for other species allows us to 
check the HVC metallicity and depletion estimates made by Lu et al. (1998).  
They derived [S/H] = $-0.60\pm^{0.11}_{0.15}$ and 
[Fe/H] = $-1.48\pm0.07$ assuming 
N(\ion{H}{1}) = $8\times10^{19}$ cm$^{-2}$, 
as determined from a comparison of single dish 21\,cm emission measurements
and an interferometric map of the 21\,cm emission with a spatial
 resolution of 1\arcmin\ 
(Wakker et al. 1999).\footnotemark\  The interferometric data reveal a compact 
concentration of
\ion{H}{1} in the direction of NGC\,3783. 

\footnotetext{Throughout this work we use the notation: 
[X/H] = log\,(N(X$^i$)/N(H$^0$)) -- log\,(X/H)$_\odot$,
where X$^i$ is the dominant ionization stage of element X in neutral gas, 
and (X/H)$_\odot$ is the solar system meteoritic (or photospheric)
abundance ratio of element 
X to H given by  Anders \& Grevesse (1989) or Grevesse \& Noels (1993) (see
Table~2).  Since many of the ions dominant in the neutral gas can also be 
present in ionized regions, all ionization corrections
refer to the amount by which [X/H] as defined above needs to be adjusted 
to account for the fraction of ion X$^i$ produced in ionized (H$^+$) gas.}

We find that the strengths of the far-UV Fe II lines (see Table~1)
are consistent with 
those of the \ion{Fe}{2} $\lambda\lambda$2374.461, 2344.214 lines 
observed by Lu et al. (1998).  
The \ion{Fe}{2} column density for the HVC remains unchanged at 
log\,N(\ion{Fe}{2}) = $13.93\pm0.05$.  Combining this result with an 
upper limit of log\,N(\ion{Fe}{3}) $<$ 13.14 (2$\sigma$) yields 
N(\ion{Fe}{3})/N(\ion{Fe}{2}) $<0.16$.  For a dilute, fully ionized gas
with properties similar to those of the warm ionized interstellar medium
ionized by O stars ($U = \langle n_\gamma/n_H \rangle \approx 6\times10^{-5}$; 
n(H$^+$)/n(H) $>$ 0.8; n$_e$ $\approx$ 0.08 cm$^{-3}$), 
N(\ion{Fe}{3})/N(\ion{Fe}{2}) $\gtrsim1$ (Sembach et al. 2000a).  Therefore, 
the expected ionized gas contribution  to the observed \ion{Fe}{2} 
column is $\lesssim10-15$\%.  Unfortunately, it is not 
possible to estimate the strength of the 
\ion{S}{3} $\lambda1012.502$ line in the HVC for comparison with the previous
\ion{S}{2} measurements.  The line is blended with strong 
high velocity H$_2$ Werner (0--0) Q(3) 1012.680\,\AA\ and Lyman (7--0) R(0) 
1012.810\,\AA\ absorptions and low velocity Lyman (7--0) R(1) 1013.435\,\AA\ 
absorption.  
Corrections accounting for the amount of singly ionized Si, P, and S 
arising in \ion{H}{2} regions
could be somewhat higher than for Fe in the same type of gas but would still
be modest.  Our \ion{Fe}{3}/\ion{Fe}{2} result confirms the simple 
ionization model advanced by Lu et al. (1998) that showed that ionization 
corrections to the abundances derived for the neutral gas from singly 
ionized species should be less than 20\%.  

Higher ionization gas does not contribute significantly to the amount of 
material in the HVC along the NGC\,3783 sight line.  Lu et al. (1994)
estimated W$_\lambda$(\ion{C}{4}\,$\lambda$1548.195) $<$ 70 m\AA\
(2$\sigma$) and 
W$_\lambda$(\ion{N}{5}\,$\lambda$1238.821) $<$ 40 m\AA\
(2$\sigma$), which correspond to log\,N(\ion{C}{4}) $<$ 13.24
and log\,N(\ion{N}{5}) $<$ 13.27.  The \ion{C}{4} result implies
that the amount of ionized hydrogen associated with highly ionized 
gas is log\,N(\ion{H}{2}) $<$ 17.3, or less than 0.2\% of the \ion{H}{1}
column, assuming a metallicity of 0.3 solar and a factor of 2 depletion
of carbon onto dust grains.
Although \ion{O}{6} absorption is observed toward NGC\,3783 at 
lower velocities (V$_{LSR}$ $<$ 150 \kms), it is not possible to  quantify
the amount of \ion{O}{6} in the HVC since it is confused by 
redshifted intrinsic \ion{H}{1} Ly$\beta$ absorption\footnotemark\ near the 
\ion{O}{6} $\lambda$1031.926 line 
and by Galactic interstellar lines (high velocity \ion{C}{2} 
$\lambda1036.337$ and H$_2$ Lyman (5-0) R(1) $\lambda1037.149$, and 
low velocity H$_2$ Lyman (5--0)
P(1) $\lambda1038.157$ and R(2) $\lambda1038.689$) near the 
\ion{O}{6} $\lambda$1037.617 line. 

\footnotetext{NGC\,3783 has a reported
redshift of $0.00973\pm0.00001$ ($2919\pm3$ \kms)
(Theureau et al. 1998).  Variable intrinsic absorption has been noted;
several prominent absorbers are present at 
velocities of $\sim-560$ and $\sim-1420$ \kms\ with respect to the systemic
velocity of the nuclear emission lines (Crenshaw et al. 1999).  
Additional
highly ionized absorbers within $\sim600$ \kms\ of the systemic velocity
have also been seen with the Chandra X-ray Observatory (Kaspi et al.
2000). }

A summary of the 
abundances for the HVC based upon the HST and FUSE data is given in Table~2.
These results are presented graphically in Figure~4.
Sulfur and phosphorus, which are elements found primarily in the gas and
are not heavily incorporated into dust grains
(Jenkins 1978; Savage \& Sembach 1996), have abundances  of 
about $0.2-0.4$ solar in the HVC; they should provide reliable 
measures of the metallicity of the gas.  The limits for N and Ar loosely 
bracket the S and P values, which is 
reassuring since N and Ar are found predominantly in neutral (\ion{H}{1})
regions.  The metallicity of the HVC is identical to the value of 
(S/H) $\sim$ 0.2--0.4
solar found for the Magellanic Stream in the direction of Fairall~9 
($l = 295.07\degr$, $b = -57.83\degr$) (Lu, Savage, \& Sembach 1994b; 
Gibson et al. 2000a). Comparing the abundances of  S and P with those of
Si and Fe shows that these latter two elements are incorporated into dust 
grains.  We find [Fe/S] = $-0.88\pm^{0.12}_{0.17}$
and [Si/S] = $-0.21\pm^{0.14}_{0.21}$.  These relative gas-phase abundances  
are similar to those determined for warm clouds in the Galactic halo 
(Sembach \& Savage 1996) and the Magellanic Bridge (Lehner et al. 2000).
The relative levels of Si and Fe depletion are intermediate to the levels
derived for gas in the Small Magellanic Cloud (SMC) and the Large Magellanic 
Cloud (LMC).  For the Sk~108 sight line in the SMC, Welty et al. (1997) find 
[Zn/H]$_{SMC}$ = $-0.64\pm^{0.13}_{0.17}$, 
[Si/Zn]$_{SMC}$ = $+0.07\pm^{0.07}_{0.09}$, 
and [Fe/Zn]$_{SMC}$ = $-0.57\pm^{0.07}_{0.09}$.  Slightly more severe
Si and Fe depletions are observed in the LMC ([Si/Zn]$_{LMC}$ $\approx$
$-0.02$ to $-0.18$;  [Fe/Zn]$_{LMC}$ $\approx$ $-0.41$ to $-1.18$;
Welty et al. 1999), but the overall metallicity of HVC\,287.5+22.5+240
appears to be at least a factor of 2 lower than that of the LMC gas.

\subsection{Molecular Hydrogen}

H$_2$ lines arising in the low velocity ISM and the HVC blanket 
the NGC\,3783 spectrum at 
wavelengths $\lambda < 1120$\,\AA.  Absorption by molecular hydrogen at the 
velocities of the 
HVC is most readily apparent in the $J = 1-3$ rotational levels in 
the Lyman series (0--0), (3--0), and (4--0) vibrational 
bands shown in Figure~2. 
The J = 0 lines in the spectrum are frequently blended with other features, 
making a determination of N(H$_2$) in the $J = 0$ level particularly 
difficult.  Most of the measurable lines in levels $J = 1-3$ 
have equivalent widths of 50--200 m\AA.  No detectable absorption is present 
in levels $J \ge4$.

To estimate the column densities of the $J = 0-3$ lines,
we measured the equivalent widths of all lines free of blending 
with other features.  These values are listed in Table~1.  We then constructed 
a single-component Maxwellian 
curve of growth and minimized the residuals about the 
best fit curve for the lines within each level.  This resulted in the 
curve of growth shown in Figure~5.  We find a Doppler parameter b $=12\pm2$
\kms\ and column densities log\,N$_J$(H$_2$) = $16.24\pm0.20$, $16.64\pm0.10$, 
$15.20\pm0.08$, and $14.80\pm0.08$ 
for $J$ = 0, 1, 2, and 3, respectively (Table~1).  The errors on N$_J$(H$_2$)
are the standard deviations of the data points about the best fit values of 
log (N$f\lambda$) for the b =12 \kms\ single-component 
curve of growth in Figure~5.  The value of N$_1$(H$_2$) and its error are more 
uncertain than for the other levels because of the small number of data 
points available.

The H$_2$ b-value is smaller than the 
b-value of 18 \kms\ found by fitting the \ion{Fe}{2} lines to a 
single-component curve of growth.  We interpret this to mean that the H$_2$
arises within a more confined region of the HVC than the \ion{Fe}{2}.  In
both cases it is likely that the broadening of the profiles is
governed mainly by turbulent motions and substructure within the 
profiles.  As long as the distribution of component widths is not strongly
bimodal, a single-component curve of growth approximation and the N$_a$(v)
analyses used in this work should yield reliable column densities. (See
Jenkins 1986 and Savage \& Sembach 1991 for discussions of
how varying component distributions can affect the column density results.)

We checked the single component curve-of-growth results by fitting 
instrumentally-smeared Gaussian profiles of varying 
column density and width to the H$_2$ lines listed in Table~1.  
We assumed an instrumental resolution of $\sim20$ \kms\ (FWHM) for all of the 
lines studied.    
Values of b smaller than 10 \kms\ do not match the data in all rotational 
levels simultaneously (i.e., not all levels fall on the same curve of growth).
Values of b larger than 15 \kms\ produce profiles that are broader than
the observed profiles.

The total H$_2$ column density of the HVC in levels $J = 0-3$
for b = 12 \kms\
is log\,N(H$_2$) = $16.80\pm0.10$. 
We find log\,N$_J$(H$_2$) $\lesssim$ 14.5 for levels 
$J \ge$ 4.  The fractional abundance of molecular hydrogen
in the 
HVC is f$_{H_2}$ = 2N(H$_2$)/[N(\ion{H}{1})+2N(H$_2$)] = $1.6\times10^{-3}$.
This value\footnote{This standard
definition of f$_{H_2}$ does not account for N(H$^+$), 
which is generally small in comparison to the amount of neutral gas present in
the interstellar clouds considered here.} falls in between the values 
 of f$_{H_2}$ $<2\times10^{-4}$ and
f$_{H_2}$ $>5\times10^{-2}$ found for 
most diffuse interstellar clouds  (Spitzer \& Jenkins 1975).
In Figure~6 we show an excitation diagram for the first five rotational
levels of H$_2$ in the HVC.  The column density of each level 
divided by the statistical
weight of the level is plotted against excitation energy.  A Boltzmann 
distribution described by a single temperature does not adequately 
describe the distribution of the data points.  We find that a temperature,
$T_{01} = 133\pm^{37}_{21}$\,K describes the relative populations of the 
$J=0$ and $J=1$
levels, while a higher temperature, $T_{23} = 241\pm^{20}_{17}$\,K, 
is needed 
to describe the relative populations of the $J=2$ and $J=3$ levels.  In
both cases, the temperatures were derived from the usual ratios of column 
densities and statistical weights (Spitzer et al. 1974) given by

\begin{equation}
T_{ij} = \frac{E_j-E_i}{k \ln[(g_j/g_i)(N_i/N_j)]}
\end{equation}

The derived value of $T_{01} \approx 133$ K 
is higher than the value of $\langle T_{01}
\rangle \sim 80$\,K derived for sight lines in the Galactic disk 
(Spitzer \& Cochran 1973; Savage et al. 1977), but is similar to the 
values found for diffuse interstellar clouds along high latitude 
sight lines toward quasars and AGNs (Shull et al. 2000b).
An elevated rotational excitation temperature derived for levels $J \ge 2$ 
compared to the 
temperature derived from levels $J = 0-1$ is common in studies of 
H$_2$ absorption and may be due to a variety of processes, including 
photon pumping of the higher $J$ levels, shocks, and
formation of H$_2$ on dust grains
(see Shull \& Beckwith 1982 and references therein).  This last 
mechanism, ``formation pumping'',  involves the excitation of rotational and
vibrational states produced when newly-formed molecules
are ejected from grain surfaces.  The 4.48 eV of H$_2$ latent 
heat of formation can appear as kinetic energy, ro-vibrational
($v,J$) excitation, or phonon (heat) energy in the grain, 
all in highly uncertain proportions.  If even 10--20\% of this energy 
appears as H$_2$ excitation, significant excitation of the upper-$J$ states 
in $v = 0$ is possible.  Radiative cascade from the initial formation entry,
through the quadrupole lines from low vibrational states, removes a good 
deal of the memory of the initial excitation.  The resulting rotational 
excitation is similar to that of radiative excitation. 

\subsection{The Intermediate Velocity Cloud Near +62 \kms}

H$_2$ absorption is convincingly detected only near 
0 and +240 \kms\ in the direction of NGC\,3783 
(see Figure~3).  The intermediate velocity gas seen 
near +62 \kms\ in \ion{H}{1} 21\,cm emission and in the \ion{Si}{2} 
and \ion{Fe}{2}
absorption lines shown in the left panel of Figure~3 is 
free of H$_2$ absorption down to a level of log\,N(H$_2$) $\lesssim$ 15.6.
Table~3 contains upper limits on N$_J$(H$_2$) derived from several transitions 
for which the 30--75 \kms\ velocity range does not contain absorption due to 
other lines. A more stringent limit of log\,N(H$_2$) $\lesssim 15.0$ 
can be derived by assuming $T_{01} > 100$\,K  and using the observed limit 
of log\,N$_1$(H$_2$) $<$ 14.74.  For this intermediate velocity cloud (IVC), 
Wakker et al. (2000) find log\,N(\ion{H}{1}) 
$\approx$ 20.0 from 21\,cm data obtained with a 15\arcmin\ beam at 
Parkes.  Using
this value, we derive f$_{H_2}$ $\lesssim$ $2\times10^{-5}$ if T $>$ 100 K.

Direct integration of the \ion{S}{2} and \ion{Fe}{2} apparent column
density profiles obtained with the GHRS and FUSE  over the velocity range
+30 to +130 \kms\ yields 
log\,N(\ion{S}{2}) = $15.01\pm0.05$ and log\,N(\ion{Fe}{2}) = $14.74\pm0.03$.  
The resulting gas-phase abundances are  
[S/H] = $-0.27\pm0.05$ (as found by Lu et al. (1994a)) and 
[Fe/S] = $-0.26\pm0.06$, which implies that most of the Fe is in the gas.
Some of the intermediate velocity gas could be ionized; the amount of 
\ion{Fe}{3} over the same velocity range is  log\,N(\ion{Fe}{3}) $<$14.22,
which implies that most ($>70$\%) of the \ion{Fe}{2} is in
the neutral gas.
The Fe depletion is very small compared to values expected for clouds in the 
Milky Way disk, where [Fe/S] $\sim$ --1.2 to --2.5 dex 
(Savage \& Sembach 1996).

The scarcity of H$_2$ and dust grains in the intermediate velocity gas 
compares favorably to recent H$_2$ measurements for other IVCs 
located in the low halo.  In the direction of 
PG\,0804+761, Richter et al. (2000) find [Fe/H] = $-0.75\pm0.05$ and 
f$_{H_2}$ = $2.97\times10^{-5}$ in the --55 \kms\ gas of the Low Latitude
Intermediate Velocity Arch (z $\approx$ 0.6--1.2 kpc).  Toward HD\,93521, 
Fitzpatrick
\& Spitzer (1993) find [Fe/H] = $-0.61\pm0.11$ in the warm halo clouds near
--60 \kms\ (z $\sim$ 0.3--1.5 kpc), and the H$_2$ measurements of Gringel 
et al. (2000) yield f$_{H_2}$ = $2.0\times10^{-5}$ assuming 
log\,N(\ion{H}{1}) = 19.61 for the same clouds (components 1--5 
in Table~2 of Spitzer \& Fitzpatrick 1993). 
Thus, the observed velocity, molecular fraction, and dust content 
of the IVC  toward NGC\,3783 are consistent 
with the idea that the IVC is 
part of a co-rotating, off-plane (z $\approx$ 5 kpc) extension 
of an outer spiral arm as suggested by Lu et al. (1994a). 

\section{Discussion}

\subsection{The Origin and Properties of HVC\,287.5+22.5+240}

The elemental abundances, kinematics, and position of HVC\,287.5+22.5+240
all indicate that the HVC is part of the Leading Arm of the Magellanic
Stream.  The tidal plus weak drag model proposed by Gardiner (1999) shows
that the material in the Leading Arm
 was stripped primarily from
the disk of the SMC and is located a distance of about 50--75 kpc from the Sun.
This feature is observed in sensitive \ion{H}{1} 21\,cm maps of the 
Magellanic Stream (Putman et al. 1998, 2000).  A detailed
study of the high velocity \ion{H}{1} 21\,cm emission in the direction 
of NGC\,3783 at different angular resolutions by Wakker et al. (1999) reveals
that discrete structures exist on scales down to 1\arcmin\
(= 14.5 pc at a distance  D = 50 kpc).  They estimate that the HVC is part
of an ensemble of small gas clouds with internal thermal 
pressures, $P/k \sim 
18,000$\,D$^{-1}_{kpc}$\,K\,cm$^{-3}$, implying $P/k = 1.1\,n_H\,T 
\sim360$ K cm$^{-3}$ for 
an assumed distance of 50 kpc.  

The number of high velocity clouds for which both gas-phase abundances and
molecular content are available is very limited.  The two highest
quality measurements to date are for HVC\,287.5+22.5+240 (the subject of 
this paper) and Complex~C.  For HVC\,287.5+22.5+240, we find 
N(H$_2$) = $6.3\times10^{16}$ cm$^{-2}$ 
and f$_{H_2}$ = $1.6\times10^{-3}$.  We 
infer the presence of dust from the relative abundances of Si, S, and Fe
(\S\,4.1).
For Complex~C in the direction of Mrk~876 ($l=98.27\degr, b = +40.38\degr$),
Murphy et al. 
(2000) find N(H$_2$) $<2\times10^{14}$
cm$^{-2}$ and f$_{H_2}$ $< 1.7\times10^{-5}$.  Combined with the abundance
information available for Fe from FUSE (Murphy et al. 2000) and 
sulfur from HST (Wakker et al. 1999; Gibson et al. 2000b), it appears that 
Complex~C 
has a low metallicity ($\sim0.1-0.3$ solar) and essentially no depletion of 
Fe onto dust grains.  Thus, these two cases imply that 
dust and H$_2$ are related in HVCs -- low values of f$_{H_2}$ and mild
depletions appear to go together.

High velocity H$_2$ has also been reported for HVCs in the direction of the 
LMC, although the detections are not as secure as the detection for
HVC\,287.5+22.5+240. Using the echelle spectrograph on 
the {\it Orbiting and Retrievable Far and Extreme 
Ultraviolet Spectrometers} (ORFEUS), Richter et al. (1999) found 
N(H$_2$) $=(2.2-3.6)\times10^{15}$ cm$^{-2}$ and 
f$_{H_2}$ $=(3.6-6.0)\times10^{-4}$ in the +120 \kms\ HVC 
toward the LMC 
star HD\,269546 ($l = 279.32\degr, b = -32.77\degr$)\footnote{As this paper
was being refereed, HD\,269546 (Sk\,-68\,82) was observed with FUSE.  The
FUSE spectra indicate that the H$_2$ 
absorption is weaker than deduced from the ORFEUS data (P. Richter, private 
communication).  This suggests that f$_{H_2}$ for the +120 \kms\ gas
in the LMC direction may be less than the value stated above.}.  
Their analysis of 
{\it International Ultraviolet Explorer} (IUE) \ion{Fe}{2} data 
indicated that (Fe/H) in the cloud is roughly half the solar value.  
Several higher quality observations of other LMC stars by FUSE also
reveal H$_2$ absorption and will allow a more thorough investigation of the 
composition and molecular fraction of the high velocity gas in front of 
the LMC 
(Richter et al. 2001).

\subsection{Where did the H$_2$ Originate?}

The presence of molecular gas in HVC\,287.5+22.5+240 requires that either the
H$_2$ formed in situ or that the H$_2$ was originally formed in the SMC 
and survived the tidal stripping that created the 
Leading Arm.    For the six LMC and one SMC
sight lines for which Magellanic Cloud 
H$_2$ was observed with ORFEUS or the {\it Hopkins
Ultraviolet Telescope} (HUT), four have values of f$_{H_2}$ $< 1\times10^{-4}$,
one has a value of $6.6\times10^{-3}$, and two have 
values $> 3.8\times10^{-2}$  (see Richter 2000 and references therein).
FUSE observations of a much larger sample of stars ($\sim40$)  
demonstrate that  H$_2$ is often present in the Clouds; preliminary results 
for 10 stars in both the LMC and SMC (Tumlinson et al. 2001)
yield $\langle$f$_{H_2}$$\rangle$ $\approx$ $1.5\times10^{-2}$ in the LMC and 
$\langle$f$_{H_2}$$\rangle$ $\approx$ $9.0\times10^{-3}$ in the SMC, with 
a range of values encompassing both the low and high f$_{H_2}$ branches.   
However, few points fall in the ``transition''
region between low and high f$_{H_2}$.  In this respect, the Magellanic 
Cloud molecular hydrogen fractions resemble those found for Milky Way 
disk and halo gas
(Savage et al. 1977; Shull et al. 2000b).
If the H$_2$ in HVC\,287.5+22.5+240 is remnant material 
stripped from the SMC, it likely originated in a region having a higher 
value of f$_{H_2}$ than the presently observed value of $1.6\times10^{-3}$.

The formation and destruction of H$_2$ are influenced by many 
competing factors.  These
include the quantity and type of dust grains, the 
gas density, the amount of self-shielding by the H$_2$, the intensity of the 
far-ultraviolet radiation field, the presence of hot gas, and the
effects of shocks. At the
column density levels observed, it is unlikely that the H$_2$ in the 
HVC is thermalized, and in a dynamic environment such as the Magellanic Stream
the competing production and destruction effects may not be equilibrated.
Thus, it is difficult to quantify the net H$_2$ production rate.  However,
it is still useful to briefly 
consider what some of the competing effects may be.

Dust grains serve as catalysts for the formation of H$_2$
(Hollenbach, Werner, \& Salpeter 1971; Spitzer \& Jenkins 1975) 
and also act as a shielding agent
against ultraviolet photons capable of dissociating the molecules.  
The formation of H$_2$ proceeds most readily for high grain surface 
area and cold gas and grain temperatures.
The heavy element abundance pattern in the HVC is similar to that in warm 
(T $\sim 10^3$ K) clouds located in the Galactic halo, which
suggests that the grains have been stripped of their outer mantles and
consist mainly of their residual (oxide) cores (Sembach \& Savage 1996).  
Thus, much of the grain surface has probably been modified and may be 
less conducive to H$_2$ formation than the surfaces of 
grains found in cold molecular clouds in the Galactic disk.  

The ultraviolet radiation field at the location of the Leading Arm is
probably less intense than for typical interstellar
environments within either the Galaxy or the SMC.  The exact details
of the radiation field and heating of the gas 
depend upon the transfer of photons out of the 
Galaxy and the Magellanic Clouds (see Wolfire et al. 1995;
Bland-Hawthorn \& Maloney 1999).      The 
rotational excitation temperatures derived for HVC\,287.5+22.5+240
are consistent with modest photon pumping of the $J = 2$ and 3 levels
but are not indicative of very high levels of excitation like those 
found for clouds near hot stars (no higher rotational level lines are
seen).  To date, no early-type stars
have been found in the Magellanic Stream.
In the Galaxy, much higher excitation temperatures ($T_{ex} > 500$ K)
are often found for H$_2$ in the vicinity of hot stars, and similarly
high values have been observed for interstellar gas in the Magellanic Clouds
(e.g., Friedman et al. 2000; Mallouris et al. 2001; Tumlinson et al. 2001).  

The absorbing clouds have 21\,cm emission structure on the 
scale of $\sim 1\arcmin - 10\arcmin$ (Putman \& Gibson 1999;
Wakker et al. 1999), which corresponds to sizes of $\sim 1-100$ pc 
at the estimated 50 kpc distance of the HVC.   
The corresponding temperature range is $\sim 100 - 1000$\,K, if the 
pressure model of \S5.1 is correct.  Directly along the NGC\,3783 sight line,
the \ion{H}{1} 21\,cm emission has FWHM $\approx$ 22 \kms\ (see Figure~2
in Lu et al. 1998).  For a line broadened solely by thermal gas motions, 
this corresponds to $T = 21.7 (FWHM)^2 \approx 10^4$ K.  However, this is an 
upper limit since the line may also be broadened significantly by turbulence
and sub-component structure.  This appears to be the case since the lines 
of species heavier than \ion{H}{1} (e.g., \ion{Si}{2}, \ion{Fe}{2}) are 
also broad and have effective 
b-values much larger than those expected for thermal 
motions alone (see Figure~3 and \S\,4).

If we adopt a gas pressure, $P/k = 360~{\rm cm}^{-3}$~K, 
and assume $T = (100~{\rm K})T_{100}$ with He/H = 0.1, the 
hydrogen density is $n_H = (3.3~{\rm cm}^{-3})(T_{100})^{-1}$, 
from which we estimate a characteristic cloud dimension 
of $L = N({\rm H~I})/n_H = (8~{\rm pc})(T_{100})$.   
The observations are therefore consistent with 
$0.3 < n_H < 3$ cm$^{-3}$ for $100 < T < 1000$~K.  
This range of densities is in the low-$n_H$ limit for models of
H$_2$ radiative excitation and collisional de-excitation.  
Preliminary estimates based on fitting the populations of rotational levels
$J = 2$ and $J=3$ suggest that the H$_2$ absorption rate in
the Lyman and Werner bands (1000 \AA) is  
$\beta_{\rm uv} < 5 \times 10^{-11}$ s$^{-1}$, 
or $<0.1$ times the average value, 
$\beta_0 = 5 \times 10^{-10}$ s$^{-1}$ in the solar 
neighborhood (Jura 1974).  

The other general statement that can made concerning the
origin of H$_2$ in this HVC is the characteristic formation time.  
The H$_2$ formation rate depends on the collision rate between 
\ion{H}{1} atoms and grains, the probability that \ion{H}{1} 
atoms are adsorbed 
on the grain surface (``sticking probability''), the mobility and 
lifetime of the atoms on the surface, and the probability that a 
molecule is ejected after formation.    
The volume formation rate of H$_2$ on grain surfaces 
(Hollenbach \& Salpeter 1971; Shull \& Beckwith 1982) is written
as $R n_H$, where $R \approx (1-3) \times 10^{-17}$
cm$^3$~s$^{-1}$ in the local ISM.  Molecule formation proceeds
rapidly for grains with surface temperatures less than a
critical value, $T_{\rm cr} \approx 20-100$ K;  H$_2$ formation 
may also be suppressed when $T_{\rm HI} \gg 100$~K.    
For the range of physical conditions for this HVC, we estimate the
H$_2$ formation time to be,
\begin{equation}
   t_{\rm form} \approx \left[ R n_{\rm HI} \right] ^{-1}
              \approx  1 \times 10^8 ~{\rm yr} 
\end{equation}
for $R \approx 10^{-17}$ cm$^3$~s$^{-1}$ and $n_H = n_{\rm HI} 
\approx 3$ cm$^{-3}$.   
Although this time is a significant fraction of the ($\sim1$ Gyr)
orbital period of the Magellanic Stream, new H$_2$ could form
up to the observed fraction, $f_{\rm H2} \approx 10^{-3}$,
on the surfaces of dust grains that survived the process of
tidal stripping. More likely, much of the observed H$_2$
survived stripping and has resisted photodestruction
by the high-latitude ultraviolet radiation field because of self-shielding
in the dissociating Lyman bands.  Self-shielding becomes important
and reduces the destruction rate (\,$\langle k \rangle \beta_{\rm uv}$\,) by
a substantial factor once N(H$_2$) $\gg$ 10$^{14}$ cm$^{-2}$ (optical
depth unity in the Lyman band).  

\subsection{Relevance to QSO Absorption Lines}
Absorption line spectroscopy of the Lyman and Werner bands of H$_2$
provides the most sensitive means available for assessing the molecular
content of cold gas in the Universe. To date, detections of H$_2$ 
absorption outside the Milky Way have been limited to the Magellanic 
Clouds (see Richter 2000; Shull et al. 2000b; Mallouris et al. 2001;
Tumlinson et al. 2001), a few 
high redshift ($z \gtrsim 2$) quasar
absorption line systems (Levshakov \& Varshalovich 1985;
Ge \& Bechtold 1997; Levshakov et al. 2000), and now 
HVC\,287+22.5+240.  Although a few low column density 
quasar absorption line systems
have modest values of f$_{H_2}$, most have very little molecular gas 
and do not appear to sustain a large quantity of molecules due to the 
intense radiation field impinging on the clouds
(see Black, Chaffe, \& Foltz 1987; Levshakov et al. 2000).
These systems also contain little dust.
HVCs represent the best low-redshift analogs to the QSO absorption line
systems with 17 $\lesssim$ log(\ion{H}{1}) $\lesssim$ 19.  Therefore, a 
better understanding of the processes that affect
the elemental abundances, molecular content, and dust composition of HVCs
could provide important insights into similar quantities derived for the 
quasar absorption line systems.  This is especially important for
HVCs that originate either from tidal interactions (as appears to be the 
case for HVC\,287.5+22.5+240) or are part of a larger ensemble of 
extragalactic clouds in the Local Group that may be interacting with the 
Milky Way (see, e.g., Blitz et al. 1999; Sembach et al. 1999, 2000b).  
If such dust-bearing clouds commonly exist around other galaxies, 
particularly those at high redshift, the dust could produce intragroup 
reddening that might affect studies of objects in those distant 
galaxies (e.g., the dimming of very distant supernovae used to study the 
expansion of the Universe).

\smallskip
We thank the FUSE Science and Operations Teams for their 
dedicated efforts to make the observations described in this 
paper possible. We also thank Mary Putman and Brad Gibson for providing a 
preliminary electronic version of Figure 1, Philipp Richter and 
Daniel Welty for useful comments on the manuscript, and  Matthew Browning
for assistance with the calculation of $\beta_{\rm uv}$.
This work is based on data obtained for the FUSE Science Team by the
NASA-CNES-CSA FUSE mission operated by the Johns Hopkins University.
Financial support has been provided by NASA contract NAS5-32985.  KRS 
and JCH acknowledge partial support from NASA Long Term Space Astrophysics 
grant NAG5-3485.  JMS acknowledges support from NASA LTSA grant NAG5-7262.

\clearpage
\newpage

\begin{deluxetable}{lcccc}
\tablenum{1}
\tablecolumns{5}
\tablewidth{318pt} 
\tablecaption{HVC Atomic and Molecular Absorption Lines}
\tablehead{Atom/Ion & Wavelength\tablenotemark{a} & $f$\tablenotemark{b} & W$_\lambda\tablenotemark{c}$ & log\,N\tablenotemark{d} \\
	       & (\AA) &  & (m\AA)     & (cm$^{-2}$)}
\startdata
\ion{H}{1}  & 21\,cm & \nodata & \nodata & $19.90\pm0.05$\tablenotemark{e} \\
\\
\ion{N}{1}  & 1134.980 & $4.35\times10^{-2}$ & $84\pm18$ & $>14.52$ \\
\\
\ion{N}{2}  & 1083.990 & $1.03\times10^{-1}$ & $99\pm42$\tablenotemark{f,g} & $13.97\pm^{0.15}_{0.24}$\tablenotemark{f,g} \\
\\
\ion{P}{2}  & 1152.818 & $2.45\times10^{-1}$ & $<32$ & $<13.05$ \\
\\
\ion{P}{3}  & \phn998.000 & $1.12\times10^{-1}$ & $<40$ & $<13.61$ \\
\\
\ion{Si}{2} & 1020.699 & $1.64\times10^{-2}$ & $56\pm14$ & $14.64\pm^{0.10}_{0.13}$ \\
\\
\ion{Ar}{1} & 1048.220 & $2.44\times10^{-1}$ & $<85$ & $<14.25$\tablenotemark{h} \\
\\
\ion{Fe}{2} & 1144.938 & $1.06\times10^{-1}$ & $78\pm17$ & $13.90\pm^{0.08}_{0.10}$ \\
	    & 1143.226 & $1.77\times10^{-2}$ & $<26$ & $<14.09$ \\
	    & 1125.448 & $1.56\times10^{-2}$ & $<24$ & $<14.16$ \\
\\
\ion{Fe}{3} & 1122.526 & $1.62\times10^{-2}$ & $<24$ & $<13.14$\\
\cutinhead{H$_2$ ($J = 0$)}
0--0 R(0)   & 1108.127 & $1.66\times10^{-3}$ & $117\pm15$ & $16.24\pm0.20$\\
1--0 R(0)   & 1092.195 & $5.90\times10^{-3}$ & $85\pm31$\tablenotemark{f} & \nodata \\
2--0 R(0)   & 1077.140 & $1.15\times10^{-2}$ & $<360$\tablenotemark{i} & \nodata \\
\cutinhead{H$_2$ ($J = 1$)}
0--0 R(1)   & 1108.633 & $1.08\times10^{-3}$ & $113\pm15$ & $16.64\pm0.10$\\
0--0 P(1)   & 1110.063 & $5.73\times10^{-4}$ & $<120$\tablenotemark{j}\\
1--0 R(1)   & 1092.732 & $3.88\times10^{-3}$ & $159\pm32$\tablenotemark{f} & \nodata \\
1--0 P(1)   & 1094.052 & $1.97\times10^{-3}$ & $112\pm34$\tablenotemark{f} & \nodata \\
2--0 P(1)   & 1078.923 & $3.90\times10^{-3}$ & $165\pm15$ & \nodata\\
2--0 R(1)   & 1077.697 & $7.84\times10^{-3}$ & $201\pm20$ & \nodata \\
3--0 P(1)   & 1064.606 & $5.66\times10^{-3}$ & $188\pm14$ & \nodata \\
4--0 P(1)   & 1051.031 & $7.73\times10^{-3}$ & $170\pm15$ & \nodata \\
\cutinhead{H$_2$ ($J = 2$)} 
0--0 Q(2)\tablenotemark{k}   & 1010.938 & $2.45\times10^{-2}$ & $106\pm15$ & $15.20\pm0.08$\\ 
2--0 R(2)   & 1079.226 & $6.85\times10^{-3}$ & $81\pm14$ & \nodata \\ 
3--0 R(2)   & 1064.994 & $1.07\times10^{-2}$ & $93\pm10$ & \nodata \\ 
3--0 P(2)   & 1066.899 & $7.07\times10^{-3}$ & $80\pm12$ & \nodata \\ 
4--0 R(2)   & 1051.498 & $1.39\times10^{-2}$ & $99\pm11$ & \nodata \\ 
4--0 P(2)   & 1053.283 & $8.98\times10^{-3}$ & $93\pm10$ & \nodata \\ 
7--0 R(2)   & 1014.977 & $1.90\times10^{-2}$ & $75\pm13$ & \nodata \\ 
8--0 R(2)   & 1003.984 & $1.67\times10^{-3}$ & $98\pm22$ & \nodata \\ 
\cutinhead{H$_2$ ($J = 3$)} 
0--0 P(3)   & 1115.895 & $7.73\times10^{-4}$ & $<24$ & $14.80\pm0.08$\\ 
1--0 R(3)   & 1096.725 & $3.01\times10^{-3}$ & $<30$ & \nodata \\ 
1--0 P(3)   & 1099.786 & $2.46\times10^{-3}$ & $<30$ & \nodata \\ 
3--0 R(3)   & 1067.474 & $1.01\times10^{-2}$ & $58\pm10$ & \nodata \\ 
3--0 P(3)   & 1070.138 & $7.48\times10^{-3}$ & $51\pm10$ & \nodata \\ 
4--0 R(3)   & 1053.977 & $1.33\times10^{-2}$ & $54\pm10$ & \nodata \\ 
4--0 P(3)   & 1056.473 & $9.56\times10^{-3}$ & $<120$\tablenotemark{j} 
& \nodata \\ 
7--0 R(3)   & 1017.423 & $1.84\times10^{-2}$ & $60\pm10$ & \nodata \\   
6--0 R(3)   & 1028.983 & $1.73\times10^{-2}$ & $57\pm10$ & \nodata \\   
7--0 P(3)   & 1019.502 & $1.05\times10^{-2}$ & $45\pm10$ & \nodata \\   
8--0 R(3)   & 1006.413 & $1.58\times10^{-2}$ & $52\pm17$ & \nodata \\ 
\cutinhead{H$_2$ ($J = 4$)}
3--0 R(4)   & 1070.899 & $9.67\times10^{-3}$ & $<36$ & $<14.6$\\ 
3--0 P(4)   & 1074.314 & $7.81\times10^{-3}$ & $<28$ & $<14.5$\\ 
4--0 R(4)   & 1057.376 & $1.29\times10^{-2}$ & $<30$ & $<14.4$\\ 
4--0 P(4)   & 1060.580 & $9.84\times10^{-3}$ & $<28$ & $<14.5$\\ 
\enddata
\tablenotetext{a}{Vacuum atomic wavelength from Morton (2001) or 
molecular
wavelength from Abgrall et al. (1993a,b).  All molecular lines are Lyman
series transitions, except for the Werner band 0--0 Q(2) transition 
at 1010.939 \AA.}
\tablenotetext{b}{Atomic $f$-value from Morton (2001) except for the 
\ion{Fe}{2} lines,
which are from Howk et al. (2000).  H$_2$ $f$-values were calculated from
the emission probabilities given by Abgrall et al. (1993a,b).}
\tablenotetext{c}{Equivalent width.  Errors are 1$\sigma$ estimates that 
account for continuum placement uncertainties and statistical noise.  
Limits are 2$\sigma$ estimates.}
\tablenotetext{d}{Adopted column density.  For the atomic lines, the results 
of direct integrations of the apparent column density profiles are listed.  
For H$_2$, the column densities of the rotational levels are given assuming 
a single component Doppler-broadened curve-of-growth with b = 12 \kms.  
Limits are 2$\sigma$ estimates.}
\tablenotetext{e}{\ion{H}{1} column density from Lu et al. (1998).  This value
is based on an interferometric map with an angular resolution of 1\arcmin.}
\tablenotetext{f}{Value measured using lower quality SiC2 data because the 
line falls in the wavelength gap between LiF1A and LiF1B.}
\tablenotetext{g}{This line may contain a small amount of \ion{N}{2}$^*$
$\lambda1084.580$ associated with the IVC at +62 \kms.  The velocity difference
between the HVC and this IVC absorption is $\sim15$ \kms.  Thus, this tentative
detection of N(\ion{N}{2}) in the HVC should probably be considered an upper limit.}
\tablenotetext{h}{Column density limit assuming the equivalent width limit 
listed and a single component curve-of-growth having b = 18 \kms.}
\tablenotetext{i}{Line appears stronger than expected due to unknown source 
of blending.}
\tablenotetext{j}{Upper limit due to blending with other H$_2$ lines.}
\tablenotetext{k}{Werner series line.}
\end{deluxetable}

\clearpage
\newpage
\begin{deluxetable}{ccc}
\tablenum{2}
\tablecolumns{3}
\tablewidth{205pt} 
\tablecaption{HVC Abundances}
\tablehead{Element & (X/H)$_\odot$\tablenotemark{a} & [X/H]\tablenotemark{b}}
\startdata
N  & $8.55\pm0.05$ & $>-1.93$\\
Si & $7.55\pm0.02$ & $-0.81\pm^{0.09}_{0.12}$ \\
P  & $5.57\pm0.04$ & $<-0.42$ \\
S  & $7.27\pm0.05$ & $-0.60\pm^{0.11}_{0.15}$\tablenotemark{c}\\
Ar & $6.56\pm0.10$ & $<-0.21$ \\
Fe & $7.51\pm0.01$ & $-1.48\pm0.07$\tablenotemark{d} \\
\enddata
\tablenotetext{a}{Solar System abundance on a logarithmic scale where
the abundance of H is 12.  Meteoritic values for
Si, P, S, Ar, and Fe are from Anders \& Grevesse (1989).
The value for N is a solar photospheric value from Grevesse \& Noels (1993).}
\tablenotetext{b}{[X/H]\,=\,log\,N(X$^i$)/N(H$^0$)\,--\,log\,(X/H)$_\odot$,
where X$^i$ = \ion{N}{1}, \ion{Si}{2}, \ion{P}{2}, \ion{S}{2}, or \ion{Fe}{2}.
Errors in this quantity reflect measurement errors only and do
not include errors on (X/H)$_\odot$.  Ionization corrections
for the singly ionized species are less than 20\% (i.e., $<0.1$ dex).}
\tablenotetext{c}{Value from Lu et al. (1998).}
\tablenotetext{d}{Value from Lu et al. (1998). Using the value of 
N(\ion{Fe}{2}) = 13.90 in Table~1 for the $\lambda1144.94$ line, we find 
[Fe/H] = $-1.51\pm^{0.09}_{0.12}$,
which is consistent with the higher precision Lu et al. result.}
\end{deluxetable}

\clearpage
\newpage
\begin{deluxetable}{lcccc}
\tablenum{3}
\tablecolumns{5}
\tablewidth{0pt} 
\tablecaption{Limits on H$_2$ in the +62 \kms\ IVC}
\tablehead{ H$_2$ Line& $\lambda$\tablenotemark & $f$\tablenotemark 
& W$_\lambda\tablenotemark{a}$ & log\,N$_J$\tablenotemark{b} \\
	       & (\AA) &  & (m\AA)     & (cm$^{-2}$)}
\startdata
0--0 R(0)   & 1108.127 & $1.66\times10^{-3}$ & $<49$ 
& $<15.44\tablenotemark{c}$ \\
4--0 P(1)   & 1051.031 & $7.73\times10^{-3}$ & $<42$ & $<14.74$ \\
0--0 Q(2)   & 1010.938 & $2.45\times10^{-2}$ & $<41$ & $<14.27$ \\ 
6--0 R(3)   & 1028.983 & $1.73\times10^{-2}$ & $<34$ & $<14.32$ \\   
4--0 R(4)   & 1057.376 & $1.29\times10^{-2}$ & $<24$ & $<14.27$ \\ 
\enddata
\tablenotetext{a}{Equivalent width limit (2$\sigma$) between +30 and +75 \kms.}
\tablenotetext{b}{Column density limit (2$\sigma$) assuming a linear 
curve of growth.} 
\tablenotetext{c}{A more stringent limit of log N$_0$(H$_2$) 
$\lesssim 14.5$ can be derived by requiring $T_{01} = 100 K$.  With such a 
requirement, log N(H$_2$) $\lesssim 15.0$.}
\end{deluxetable}

\clearpage
\newpage

\figcaption{An \ion{H}{1} Parkes All-Sky Survey (HIPASS) map of the 
high velocity (+170 to +400 \kms) gas in the general direction of 
NGC\,3783 (Putman \& Gibson 1999).  The data have an angular 
resolution of approximately 15.5\arcmin.
NGC\,3783, marked by a white star symbol, lies behind a large complex of gas 
believed to be the Leading Arm of the Magellanic Stream.  A higher resolution
map of this cloud can be found in Wakker et al. (1999).}

\figcaption{A portion of the FUSE LiF1A spectrum of NGC\,3783.  Most of the
absorption lines in this spectrum are due to molecular hydrogen in the 
interstellar medium of the Galaxy and the high velocity cloud 
HVC\,287.5+22.5+240.
The identifications for these lines are shown at the top of each panel, 
with the lower set of tick marks indicating absorption at the velocity
of the HVC (+240 \kms).  Additional features due to \ion{Ar}{1} and \ion{Fe}{2}
are also indicated.  The exposure
time for this observation was 37 ksec. The data are binned to a sampling
interval of $\approx 8$ \kms\ per pixel and have S/N $\approx$ 10--15 per 
20 \kms\ resolution element.}

\figcaption{Continuum normalized profiles for selected interstellar lines
in the spectrum of NGC\,3783.  The left panel contains atomic species.  
Absorption is clearly present in the atomic lines at low, intermediate,
and high velocities.  The
right panel contains various H$_2$ lines in different rotational levels.
Absorption is present at both low and high velocities.
The +240 \kms\ velocity of the HVC is marked in each panel 
by the vertical dashed line.
When other lines are present in addition to the primary line illustrated,
the identifications (without wavelengths) are given above the spectrum.}

\figcaption{Logarithmic gas-phase abundances for Ar, P, S, Si, and Fe in  
HVC\,287.5+22.5+240.  The abundances have been normalized against the Solar
System meteoritic values listed in Table~2.  Approximate metallicities 
([Fe/H]) derived from stellar/nebular abundances are shown for the LMC 
and SMC; values of [O/H] are about 0.12 dex lower (Russell \& Dopita 1992).
The HVC data points are compared to the values shown for 
diffuse clouds in the Galactic halo and the SMC (see text).  Note that the HVC
metallicity derived from [S/H] closely matches that of the SMC.}

\figcaption{Single-component curve of growth for the H$_2$ lines identified
in the spectrum of NGC\,3783 (see Table~1).  Data points for rotational
levels $J = 0-3$ are shown with the symbol coding and the 
values of N$_J$ indicated in the legend.  The dashed lines 
indicate the curves of growth appropriate for b-values of 10 and 14 \kms.}

\figcaption{H$_2$ column density (N$_J$) divided by statistical weight (g$_J$)
as a function of excitation energy (E$_J$) for rotational levels $J = 0-4$
in HVC\,287.5+22.5+240.  The values expected for two different
Boltzmann distributions with the temperatures indicated are also shown as 
dashed lines.  Note that there appears to be a clear difference in the 
slopes of the lines derived from the low-$J$ and high-$J$ lines.  The 
data point for $J = 4$ is a 2$\sigma$ upper limit.  All other errors are 
1$\sigma$ estimates.}

\clearpage
\newpage
\begin{figure}[ht!]
\includegraphics{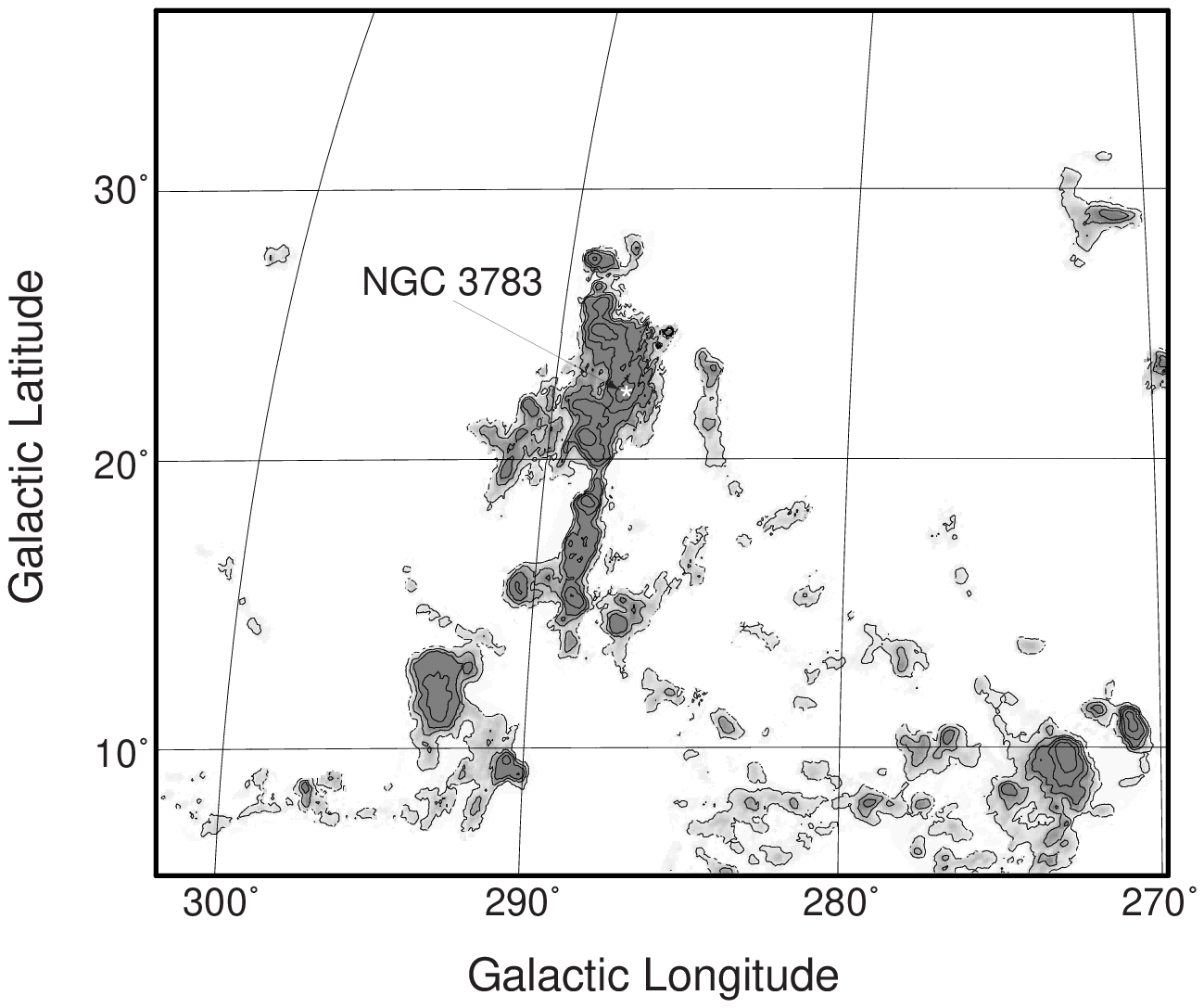}
\vspace{18.5cm}
{\huge{\bf Figure 1.}}
\end{figure}

\clearpage
\newpage
\begin{figure}[ht!]
\includegraphics{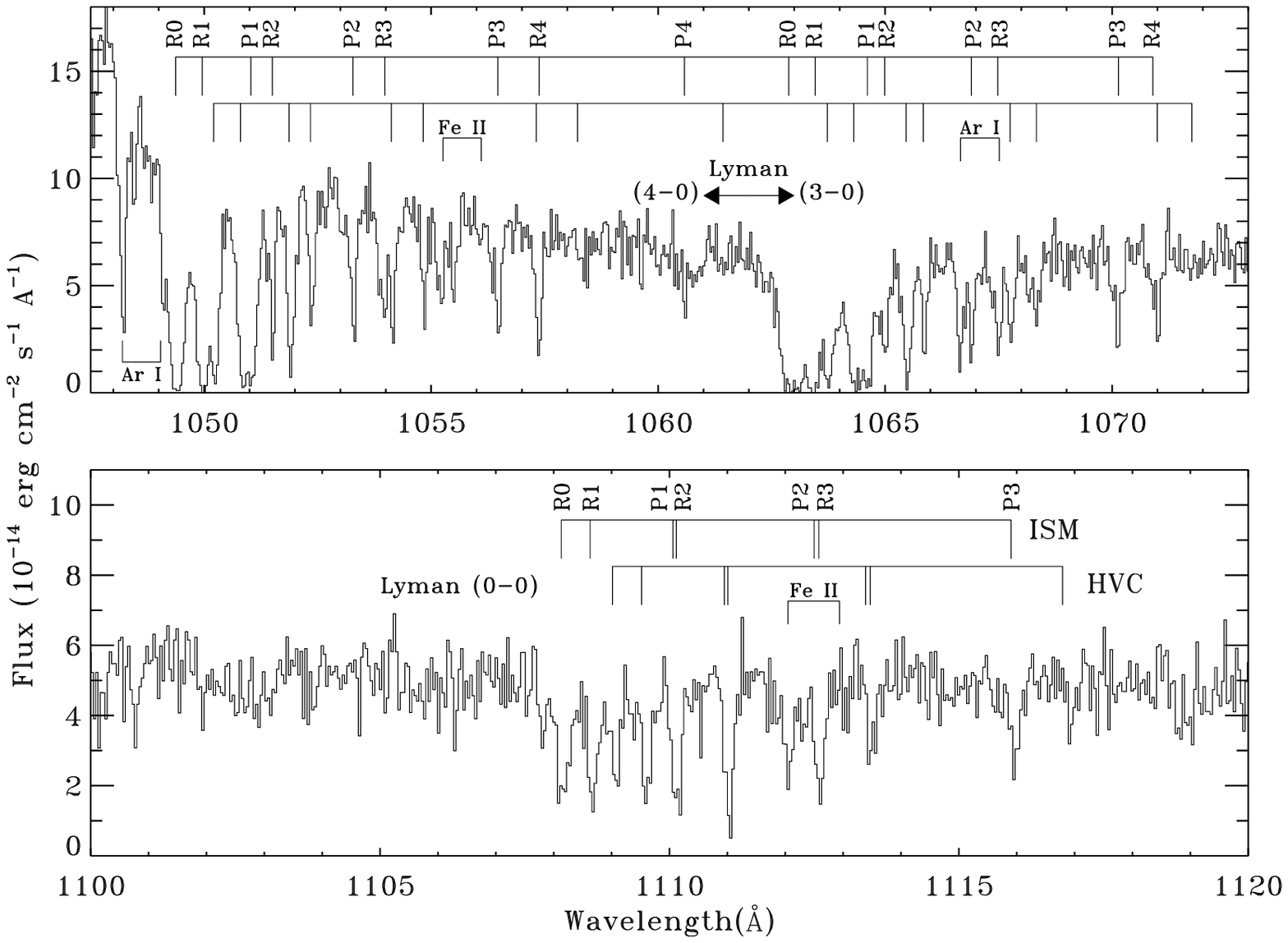}
\vspace{18.5cm}
{\huge{\bf Figure 2.}}
\end{figure}

\clearpage
\newpage
\begin{figure}[ht!]
\includegraphics{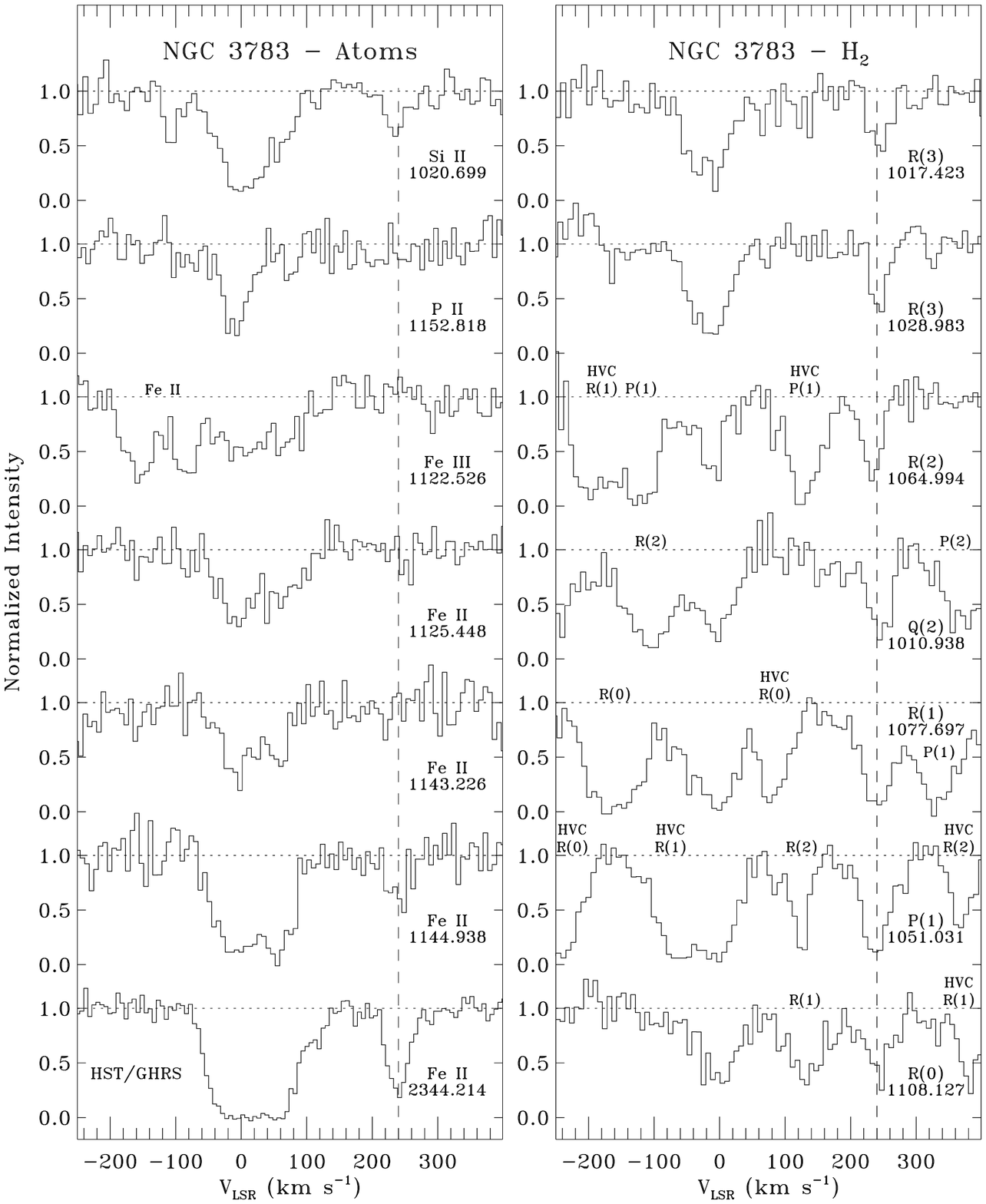}
\vspace{22.5cm}
{\huge{\bf Figure 3.}}
\end{figure}

\clearpage
\newpage
\begin{figure}[ht!]
\includegraphics{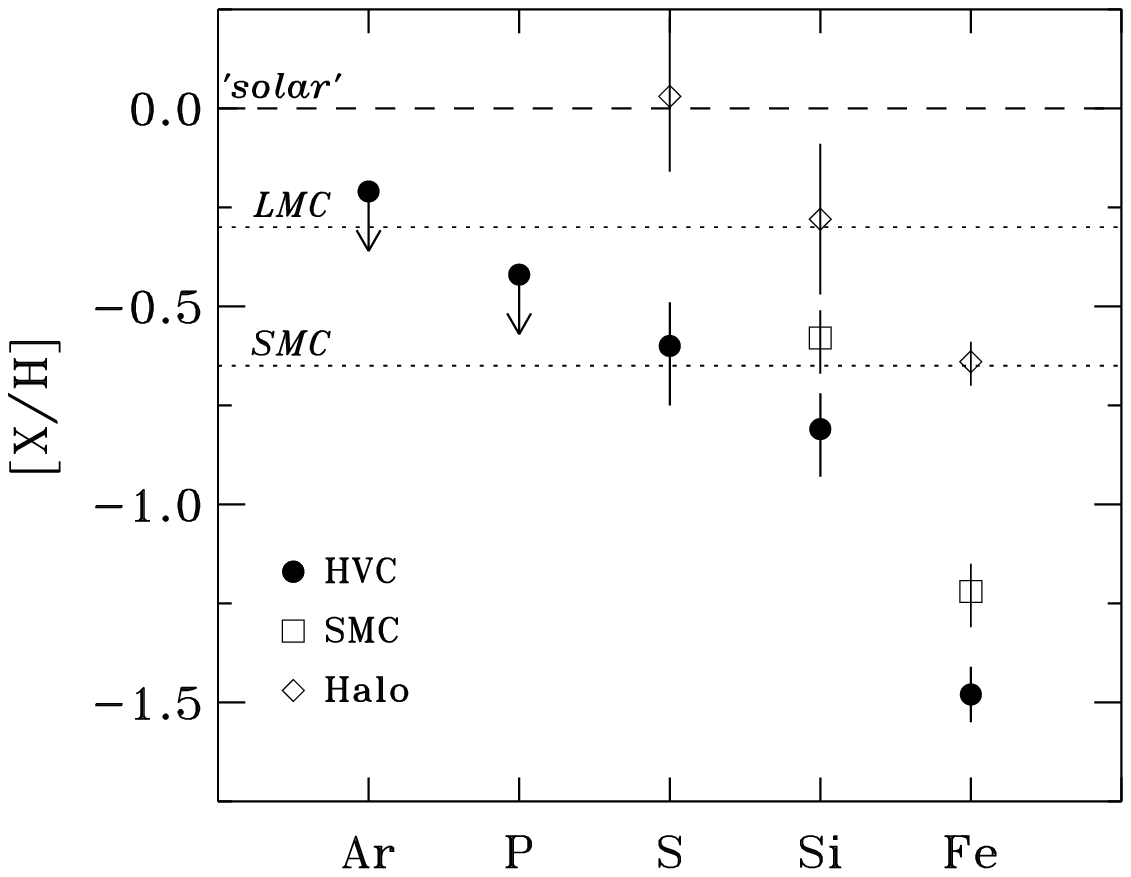}
\vspace{18.5cm}
{\huge{\bf Figure 4.}}
\end{figure}

\clearpage
\newpage
\begin{figure}[ht!]
\includegraphics{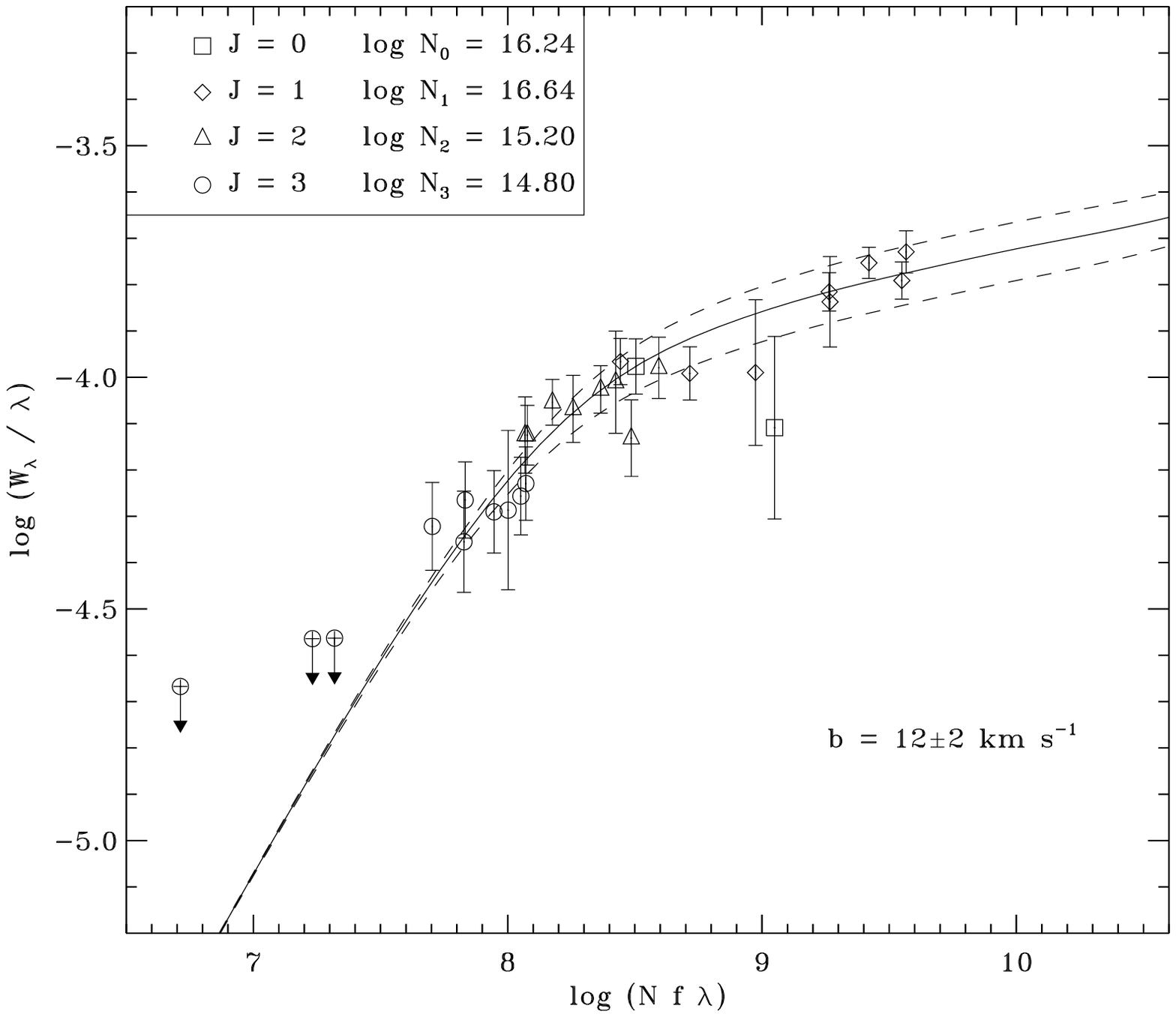}
\vspace{18.5cm}
{\huge{\bf Figure 5.}}
\end{figure}

\clearpage
\newpage
\begin{figure}[ht!]
\includegraphics{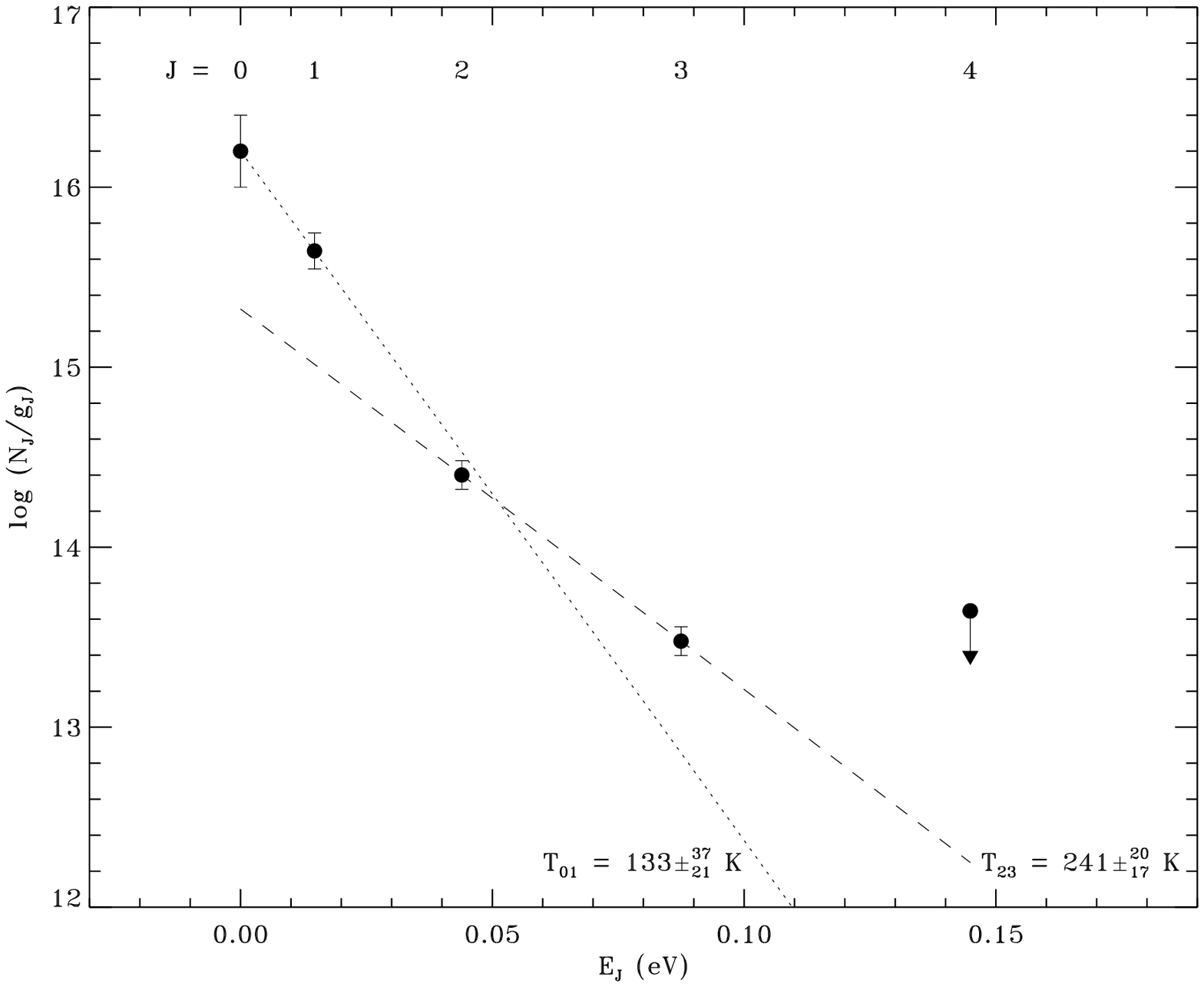}
\vspace{18.5cm}
{\huge{\bf Figure 6.}}
\end{figure}

\end{document}